\documentclass{aa}
\usepackage{graphicx}
\usepackage{color}
\usepackage{amsmath}
\usepackage{siunitx}
\usepackage{stfloats}
\usepackage{txfonts}
\usepackage[colorlinks,citecolor=blue,urlcolor=blue,filecolor=blue,linkcolor=blue]{hyperref}

\begin{document} 
\renewcommand{\sectionautorefname}[1]{Section }
\renewcommand{\subsectionautorefname}[1]{Section }

   \title{The phase-space of tailed radio galaxies in massive clusters}

   \author{S. van der Jagt \inst{1}
           \and E. Osinga \inst{1,2}
           \and R.J. van Weeren \inst{1} 
           \and G.K. Miley \inst{1}
           \and I.D. Roberts \inst{1,3,4}
           \and A. Botteon \inst{5}
           \and A. Ignesti \inst{6}}
    
   \institute{Leiden Observatory, Leiden University, PO Box 9513, NL-2333CA Leiden, The Netherlands\\
   \email{s\_vdjagt@hotmail.com}
   \and
   Dunlap Institute for Astronomy \& Astrophysics, University of Toronto, 50 St. George Street, Toronto, ON M5S 3H4, Canada
   \and
   Department of Physics \& Astronomy, University of Waterloo, Waterloo, ON N2L 3G1, Canada
   \and
   Waterloo Centre for Astrophysics, University of Waterloo, 200 University Ave W, Waterloo, ON N2L 3G1, Canada
   \and
   INAF - Istituto di Radioastronomia, via P. Gobetti 101, 40129 Bologna, Italy
   \and
   INAF - Astronomical Observatory of Padova, vicolo dell'Osservatorio 5, IT-35122 Padova, Italy}

   \date{Received 24-06-2024; accepted 26-04-2025}
 
  \abstract
  {The radio jets of radio galaxies in galaxy clusters are often bent due to the ram pressure of the intracluster medium. Most studies of bent radio tails initially identified tailed sources and then attempted to characterise their environments. In this paper we take an alternative approach, by starting with a well-defined sample of galaxy clusters and subsequently identifying tailed radio sources in these known environments. Our sample consists of 81 galaxy clusters from the \textit{Planck} ESZ cluster sample. We present a catalogue of 127 extended cluster radio sources, including brightest cluster galaxies, obtained by visually inspecting \textit{Karl G. Jansky} Very Large Array (1-2 GHz) observations. We have determined the bending angle of 109 well-structured sources, and classified them accordingly: 84 narrow-angle tailed sources (NATs), 16 wide-angle tailed sources (WATs), and 9 non-bent radio sources (i.e. with bending angles of less than \ang{15}). We find a negative correlation between the bending angle and the distance to the cluster centre (impact radius), and we observe that NATs generally have smaller impact radii than the regular galaxy population and WATs. We present a phase-space diagram of tailed radio galaxy velocities and impact radii and find that NATs have a significant excess in the high-velocity and low-impact radius region of phase space, indicating they undergo the largest amount of ram pressure bending. We compared the results from our sample with those for jellyfish galaxies, and suggest that the mechanism responsible for bending the radio tails is similar to the stripping of gas in jellyfish galaxies, although tailed radio galaxies are more concentrated in the centre of the phase space. Finally, we find that NATs and WATs have the same occurrence ratio in merging and relaxed clusters. However, their distribution in the phase-space is significantly different. We report an excess of NATs in the high-velocity and low-impact-radius phase-space region in merging clusters, and an excess of relaxed clusters in the low-velocity and low-impact-radius region.}
  
   \keywords{Galaxies: clusters -- Galaxies: jets -- Galaxies: kinematics and dynamics}

   \maketitle

\section{Introduction}
Radio sources associated with cluster galaxies often have `bent-tail' morphologies and sizes of up to $\sim$1 Mpc at low frequencies \citep[e.g.][]{2019A&A...622A..25W,2020MNRAS.493.3811S}. These extended tails are frequently observed to have non-linear structures. Interaction of the cluster radio sources with the intracluster medium (ICM) is believed to be responsible for the observed bending \citep[e.g.][]{1976ApJ...203L.107R,1979Natur.279..770B,1979ApJ...234..818J,2017PhPl...24d1402J}. Bent radio sources are unique diagnostic tools for studying galaxy clusters. Because of their large sizes, these sources provide information about the nuclear activity history of the host galaxies over hundreds of millions of years. The tails are postulated to be trails that trace the path of the host galaxy within the cluster, bent and buffeted by ram pressure exerted by the ICM \citep[e.g.][]{1972Natur.237..269M,1976ApJ...205L...1O}. The first of these tailed radio galaxies were discovered in galaxy clusters such as the Perseus, Coma, and 3C129 \citep{1968MNRAS.138....1R,1970MNRAS.151....1W,1968MNRAS.138..259M}. 

Tailed radio sources have historically been divided into two classes \citep{1980ARA&A..18..165M}: wide-angle tailed radio galaxies (WATs) and narrow-angle tailed radio galaxies (NATs). WATs (NATs) are generally defined as having bending angles smaller (larger) than 90 degrees when defining the bending angle as the deviation from a linear source  \citep[e.g.][]{2019AJ....157..126G}. At times, sources are bent extremely enough that the tails blend together. In low-resolution observations, these tailed radio galaxies seem to only have one radio jet. However, very high-resolution observations often show these objects to have two-sided jets \citep[e.g.][]{2017A&A...608A..58T,2020MNRAS.499.5791G,2021Galax...9...85R}.

Tailed radio sources are frequently found inside dense environments and are significantly more bent closer to the cluster centre, where the surrounding gas is denser \citep{2019MNRAS.488.2701M,2019AJ....157..126G}. \citet{2021MNRAS.506L..55D} find that NATs up to 7 $r_{500}$ from the cluster centre are generally on radially inbound orbits, where $r_{500}$ is the radius covering the volume of the cluster in which the mean density is 500 times the critical density of the Universe. It has been suggested that tailed radio sources have higher velocities than other galaxies in the cluster \citep{1972Natur.237..269M}, and this has been investigated for small samples of tailed radio galaxies \citep[e.g.][]{1972Natur.237..269M,1978ApJ...221..422U}. Merging galaxy clusters often have a more turbulent ICM than relaxed clusters, and therefore the dynamical state of the galaxy cluster might also affect the tailed radio galaxies. It has been suggested, based on smaller samples, that WATs are a by-product of the merging of galaxy clusters and form with the merging of clusters \citep{1994AJ....108.2031P,1995ApJ...445...80L,Gomez1997a,Gomez1997b,2000MNRAS.311..649S}. NATs are also proposed to be associated with merging galaxy clusters \citep{Bliton:1998mi}. \citet{Bliton:1998mi} studied a sample of 15 galaxy clusters with 23 NATs and found that NATs prefer dynamical complex systems.

To study how bent radio sources are influenced by the ICM and how they can be used as tracers of cluster properties, reliable samples of these objects need to be constructed. Previous studies have either used catalogues of bent radio sources to investigate whether the radio sources can be associated with nearby clusters \citep[e.g.][]{2019AJ....157..126G, 2021MNRAS.506L..55D,2022MNRAS.516..372B} or they have used tailed radio sources to find galaxy clusters \citep[e.g.][]{O'Brien:201647,Paterno_Mahler_2017}. These studies often use a combination of photometric and spectroscopic redshifts to identify clusters. A problem with these approaches is that there is still quite a large uncertainty in redshift and the false cluster association rate may be substantial, particularly in lower-mass clusters.

In this study we took a different approach, by starting from a well-defined and mass-limited sample of X-ray-defined \textit{Planck} clusters \citep[][]{2011A&A...536A...8P} and investigating tailed radio sources inside these clusters. We used our sample to investigate how the properties of tailed radio galaxies relate to the properties of their host clusters. Properties such as the mass, redshift, and dynamical state for this sample of \textit{Planck} clusters are well known from \textit{Planck} and \textit{Chandra} observations (see \autoref{sec:Data}). The dynamical states from \textit{Chandra} X-ray observations enable us to study the relation between the dynamical state of the cluster and the tailed radio galaxies further. Additionally, for a subset of clusters in our sample, we have accurate spectroscopic redshifts available for cluster members, which allow us to analyse the distribution of tails in the radial velocity-projected radial distance phase space. We also compared the tailed radio galaxies with the general cluster population and jellyfish galaxies. Jellyfish galaxies are star-forming galaxies in galaxy clusters with displaced gas due to ram pressure stripping \citep[e.g.][]{2022A&ARv..30....3B}. They are usually characterized by high velocity orbits, so that the ram pressure from the ICM is strong enough to remove their interstellar medium \citep[ISM;][]{2024ApJ...965..117B}. Jellyfish galaxies sometimes show radio continuum tails. These radio tails are not from active galactic nuclei (AGNs) but from cosmic rays from star formation that is displaced from the disc of the galaxy \citep[e.g.][]{Vollmer_2004,2020MNRAS.496.4654C,2021A&A...650A.111R,Ignesti_2022,2023A&A...675A.118I}.

In \autoref{sec:Data_Selection_Class} we present our data and describe our method for selecting and determining cluster membership and classifying the tailed radio sources. In \autoref{sec:Results} we show the distribution in phase space for tailed radio galaxies and compare it with other cluster galaxies and jellyfish galaxies. We also investigate the occurrence of tailed radio galaxies as a function of their host cluster's dynamical state. In \autoref{sec:Discussion} we discuss the results. Our conclusions are presented in \autoref{sec:Conclusion}. Throughout this study we use a flat $\Lambda$ cold dark matter cosmology with $\Omega_{M}=0.3$, $\Omega_{\Lambda}=0.7$, and H$_{0}=$70 km\, s$^{-1}$Mpc$^{-1}$.

\section{Data,selection, and classification}\label{sec:Data_Selection_Class}
\subsection{Data}\label{sec:Data}
The cluster sample we used in this work is presented in \citet[]{2022A&A...665A..71O}. The full sample consists of 124 clusters with redshifts $z$ < $0.35$ and declination > $-$\ang{40}, selected from the \textit{Planck} Sunyaev-Zeldovich (SZ) catalogues: the \textit{Planck} Early Sunyaev-Zeldovich clusters \citep[\textit{Planck}-ESZ;][]{2011A&A...536A...8P}, the \textit{Planck} catalogue of Sunyaev-Zeldovich sources \citep[PSZ1;][]{2014A&A...571A..29P} and the second \textit{Planck} catalogue of Sunyaev-Zeldovich sources \citep[PSZ2;][]{2016A&A...594A..27P}. High-quality X-ray information is available for 93 out of 124 clusters from The \textit{Chandra} \textit{Planck} Legacy Program for Massive Clusters of Galaxies\footnote{\url{https://hea-www.cfa.harvard.edu/CHANDRA_PLANCK_CLUSTERS/}}. These observations have a minimum of 10,000 source counts per cluster and comprise one of the largest matched SZ plus X-ray samples with a well-known selection function.  

These clusters have been observed by the \textit{Karl G. Jansky} Very Large Array \citep[VLA; VLA project code 15A-270][]{2022A&A...665A..71O}. The VLA observations are L-band (1–2 GHz) observations, have a resolution of approximately 3$\arcsec$, a sensitivity of 20-30 $\mu$Jy/beam for most fields, and a primary beam full width at half maximum (FWHM) of 0.5 degrees at 1.5 GHz. Because of the fixed angular size of the primary beam, this means that radio sources can be found out to larger radii in higher-redshift clusters. For clusters with redshift $z > 0.05$ ($z>0.10$) the radio image radius (FWHM) is $>r_\mathrm{500}$ ($>2r_\mathrm{500}$) \citep[cf. Fig. A.1 in ][]{Osinga2024magnetism}. Thus, to make sure we have an unbiased sampling as a function of radius, we only used clusters at $z>0.1$ and sources at radii $r<2r_\mathrm{500}$. This results in a final sample of 81 clusters, out of which 62 have \textit{Chandra} data available.

To link the tailed radio sources to their host galaxies, \citet{2022A&A...665A..71O} predominantly used the Legacy Imaging Surveys \citep{2019AJ....157..168D}. This is an optical imaging survey that mapped 14\,000 square degrees of the northern sky in the $g$,$r$, and $z$ optical bands. Because the Legacy survey does not cover the full sample, the Panoramic Survey Telescope and Rapid Response System (Pan-STARRS) imaging survey Pan-STARRS1 \citep[PS1;][]{2016arXiv161205560C} was also used. PS1 covers the northern sky above a declination of -30 degrees in the $g$, $r$, $i$, $z$, and $y$ optical bands. To confirm cluster membership and determine accurate velocities, \citet{2022A&A...665A..71O} used several redshift surveys. Spectroscopic redshifts were taken from the NASA/IPAC Extragalactic Database (NED) and the Sloan Digital Sky Survey \citep[SDSS;][]{2000AJ....120.1579Y}. We sourced additional spectroscopic redshifts from the Hectospec Cluster Survey of SZ-selected clusters \citep[HeCS-SZ;][]{2016ApJ...819...63R}. For sources where spectroscopic redshifts were unavailable, we used photometric redshifts from SDSS \citep{2016MNRAS.460.1371B}, the Legacy Survey \citep{2022MNRAS.tmp..606D}, and Pan-STARRS \citep{2020A&A...642A.102T}. Further details on the source association and redshift determination are given in \citet{2022A&A...665A..71O}.

Finally, accurate cluster centre coordinates and dynamical states were determined from \textit{Chandra} X-ray observations presented by \citet{2017ApJ...843...76A} and \citet{Andrade_Santos_2021}. If available, we followed \citet{2017ApJ...843...76A} to determine the cluster dynamical state (i.e. whether a cluster is merging or relaxed). 

To determine the cluster centre coordinates for clusters that are not covered by the X-ray observations from \citet{Andrade_Santos_2021}, we took their PSZ2 SZ-derived coordinates \citep{2016A&A...594A..27P} as the centre, which have a median offset of 0.9 arcminutes for clusters where both observations are available. The physical offset between the \textit{Planck} and \textit{Chandra} central coordinates for these clusters ranges from 15 kpc to 621 kpc, with a median offset of 190 kpc. We excluded clusters where the dynamical state is not available from analyses regarding differences between dynamical states.

\subsection{Selection}
We identified 127 extended radio sources as cluster members in 81 clusters with $z>0.1$ within 2r$_{500}$ of the cluster centre. We defined a source as being `extended' if the largest angular size is larger than 5 times the restoring beam  (15$^{\prime\prime}$). This threshold was set to generally be able to trace the morphology of the radio emission beyond the disc of the host galaxy and identify the orientation of the outflow with respect to the host galaxy \citep[e.g.][]{2020A&A...642A..70O}. We note that because of this angular size cut, we preferentially selected larger sources at higher redshifts, though these are still moderate sizes for typical radio galaxies (i.e. $\gtrsim$ 30 kpc at $z=0.1$ and $\gtrsim 75$ kpc at $z=0.35$). We find no systematic trends in terms of the classification of the source (e.g. WAT/NAT/non-bent) as a function of redshift. To ascertain whether a radio source is a cluster member its redshift ($z_{galaxy}$) was compared with the redshift of the galaxy cluster ($z_{cluster}$) as follows:

\begin{equation}\label{eq:delta_z}
    |\Delta z - \sigma| > \frac{z_{galaxy}-z_{cluster}}{1+z_{galaxy}},
\end{equation}

where $\sigma$ is the 1 sigma (68\%) uncertainty of the redshift estimate and we adopted the threshold $\Delta z = 0.04$, $c\Delta z = 11991$ km/s. This threshold is also used in the WHL catalogue \citep{2015ApJ...807..178W} to determine cluster membership and used in the \citet{2019AJ....157..126G} study of radio tails. Sources without a redshift estimate were excluded as they are likely background sources. Similarly, radio sources that do not have optically visible host galaxy were treated as background sources and not cluster members, as all galaxy clusters lie at low redshifts ($z<0.35$). In {\color{blue}\autoref{table:cluster_member}} we give an overview of the redshifts used for the cluster membership determination.

\begin{table}[b]
\centering
\caption{Overview of the number of spectroscopic and photometric redshifts used for the cluster membership determination.}
\label{table:cluster_member}
\begin{tabular}{llllll}
\hline
Sample          & NAT           & WAT           & Non bent & N.A.\tablefootmark{a}     & Total\\ \hline
NED             & 48   & 9    & 5        & 12       & 74   \\
SDSS-spec       & 1             & 1             & 0        & 2      & 4    \\
Legacy-phot     & 9    & 1    & 1        & 1        & 12   \\
PS-phot         & 26   & 5    & 3        & 3        & 37   \\ \hline
Total           & 84   & 16   & 9        & 18       & 127  \\ \hline
\end{tabular}
\tablefoot{27 out of 84 NATs and 6 out of 18 WATs also have a HeCS-SZ spectroscopic redshift. \tablefoottext{a}{N.A. refers to the extended radio sources for which no bending angle determination was possible.}}
\end{table}

\subsection{Classification}\label{sec:classification}
\begin{figure}[t]
\centering
\includegraphics[width=\hsize]{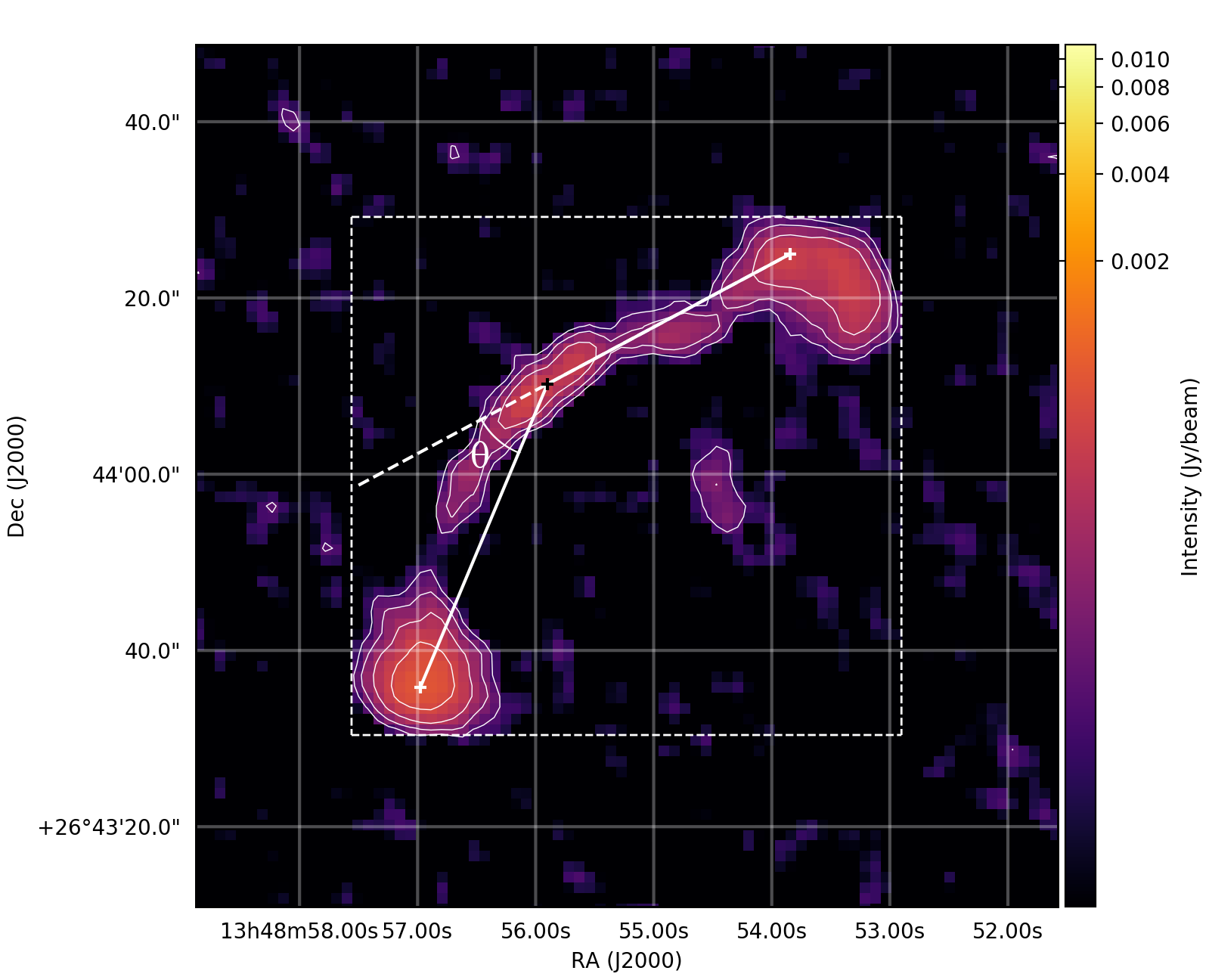}
  \caption{Example of how bending angle and size are measured. The black `$+$' shows the location of the optical counterpart, and the white `$+$' shows the location of the highest radio intensities in the lobes. The white lines connect the location with the optical counterparts with the highest intensities, which are used to calculate the bending angle $\theta$. The dashed lines show the fitted box around the edges of the 3$\sigma$ contours used to calculate the size of the source.}
     \label{fig:BA_methodology}
\end{figure}

To determine the size and bending angle of the extended radio sources, we followed the method from \citet{2019AJ....157..126G}. A rectangular-shaped box was fitted around the edges of the 3$\sigma$ contours of the radio galaxy, exemplified in {\color{blue}\autoref{fig:BA_methodology}}. The diagonal of the bounding box was taken as the largest angular size and the bending angles of the extended radio sources were calculated from the brightest spots in both lobes and the position of the optical counterpart, as shown in {\color{blue}\autoref{fig:BA_methodology}}.

We classified tailed radio galaxies into WATs and NATs based on a commonly used threshold bending angle of \ang{90} \citep[e.g.][]{ODea1985,2019MNRAS.488.2701M}\footnote{We note that \citet{2019MNRAS.488.2701M} defined the bending angle as the opening angle between the two tails, while we defined the bending angle according to \autoref{fig:BA_methodology}. However, the thresholds are essentially the same.}. Sources with a bending angle of more than \ang{90} were classified as NATs, and sources with a bending angle between \ang{15} and \ang{90} were classified as WATs. The minimum bending angle of \ang{15} was set to separate WATs from non-bent double-lobed sources. 

Most of the extended sources have clear collimated jets, but some, particularly brightest cluster galaxies, show diffuse emission with no clear structure.  We did not include any cluster galaxies without clear boundaries or linear jets. Some bent sources do not show two separate lobes, tails or jets in our observations, in which case we defined the bending angle as 180° and classified the sources as NATs. This results in a final sample of 109 sources with bending angles, of which 84 are NATs, 16 are WATs, and 9 are non-bent sources. Out of the 84 NATs, 65 have tails that are blended together and 19 are radio galaxies where both tails are still visible. To check the robustness of the classification in this work, we also performed our analysis with slightly different bending angle thresholds. We changed the bending angle threshold to separate NATs and WATs to \ang{80} and \ang{100}, and changed the bending angle to separate WATs and non-bent sources to \ang{5} and \ang{25}. No significant differences are found in our results.

\section{Results}\label{sec:Results}
\subsection{Distance to the cluster centre}
\begin{figure}[t]
\centering
\includegraphics[width=0.9\hsize]{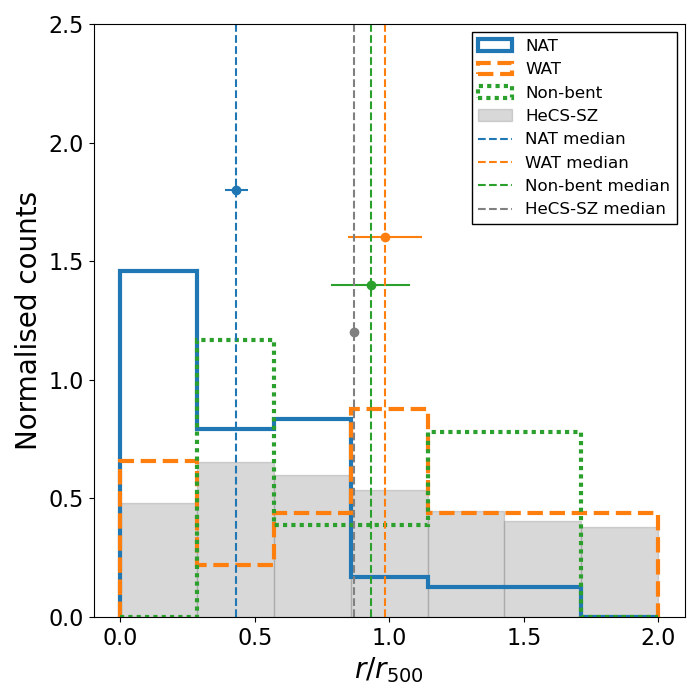}
  \caption{Distances from the cluster centre for NATs and WATs, normalised by the $r_{500}$. The dashed lines show the median values of the distribution and the error bars give the 1$\sigma$ statistical uncertainty on the median. The counts for each histogram are such that the area under the histogram integrates to 1. As a reference, the distribution of the galaxies in HeCS-SZ \citep{2016ApJ...819...63R} is given with the grey histogram.}
     \label{fig:Distances}
\end{figure}

There is evidence that tailed radio galaxies are more bent when they are located closer to the cluster centre \citep{2019MNRAS.488.2701M,2019AJ....157..126G}. This bending could be related to the increased ram pressure exerted on the galaxy by a denser intergalactic medium near the cluster centre or due to the fact that at the pericentre of their orbits, in the cluster centre, the tailed radio galaxies are moving at the highest velocity, and thus are subjected to the maximum ram pressure \citep{1979Natur.279..770B,1979ApJ...234..818J}. To investigate this relation, we calculated the projected distance between the host optical counterpart and the coordinates of the cluster centre given by \textit{Chandra} X-ray observations \citep{Andrade_Santos_2021}, or PSZ2 \citep{2016A&A...594A..27P} if \textit{Chandra} data were unavailable. The distribution of the projected distances of WATs and NATs normalised by the cluster R$_\mathrm{500}$ is illustrated in {\color{blue}\autoref{fig:Distances}}. For reference, the typical cluster galaxy population is also shown (in grey) for galaxies from the HeCS-SZ \citep{2016ApJ...819...63R}. The median values in {\color{blue}\autoref{fig:Distances}} show that the distribution of NATs is more concentrated towards the centre of the cluster than all the galaxies (>5 $\sigma$ below WATs). A two-sample Kolmogorov-Smirnov test (KS-test) to compare the NATs to the general cluster population indeed confirms that NATs are drawn from different populations than the non-tailed galaxies (p $<$ 0.001). Furthermore, the KS-test shows that NATs form a different population than the WATs (p $=$ 0.001). The KS-test confirms that NATs are intrinsically located closer to the cluster centre than the general population of galaxies and WATs.

\begin{table*}[bp]
\caption{Galaxy distribution in phase-space regions for all clusters in the HeCS-SZ sample.} 
\label{table:excess}    
\centering                      
\begin{tabular}{c c c c c c c c}     
\hline
Region  & $N^{Q_{i}}_{NAT}$ & $N^{Q_{i}}_{WAT}$ & $N^{Q_{i}}_{jellyfish}$   & $N^{Q_{i}}_{HeCS-SZ}$ & $\eta^{Q_{i}}_{NAT}$              & $\eta^{Q_{i}}_{WAT}$             & $\eta^{Q_{i}}_{jellyfish}$ \\ \hline
1       & 0                 & 0                 & 6               & 172        & 0.00$^{+0.76}_{-0.00}$   & 0.00$^{+2.85}_{-0.00}$  & 1.19$^{+0.66}_{-0.33}$                  \\
2       & 5        & 1        & 28               & 367          & 1.63$^{+0.85}_{-0.47}$   & 1.47$^{+2.06}_{-0.56}$  & 2.60$^{+0.46}_{-0.40}$                  \\
3       & 19       & 3        & 45               & 1488         & 1.53$^{+0.16}_{-0.22}$   & 1.09$^{+0.38}_{-0.38}$  & 1.03$^{+0.11}_{-0.11}$\\
4       & 3        & 2        & 16               & 1211         & 0.30$^{+0.24}_{-0.09}$   & 0.89$^{+0.57}_{-0.34}$  & 0.45$^{+0.12}_{-0.09}$                  \\ \hline
Total   & 27       & 6        & 95               & 3238       &                        &\\ \hline
\end{tabular}
\end{table*}

\begin{figure}[t]
\centering
\includegraphics[width=1.05\hsize]{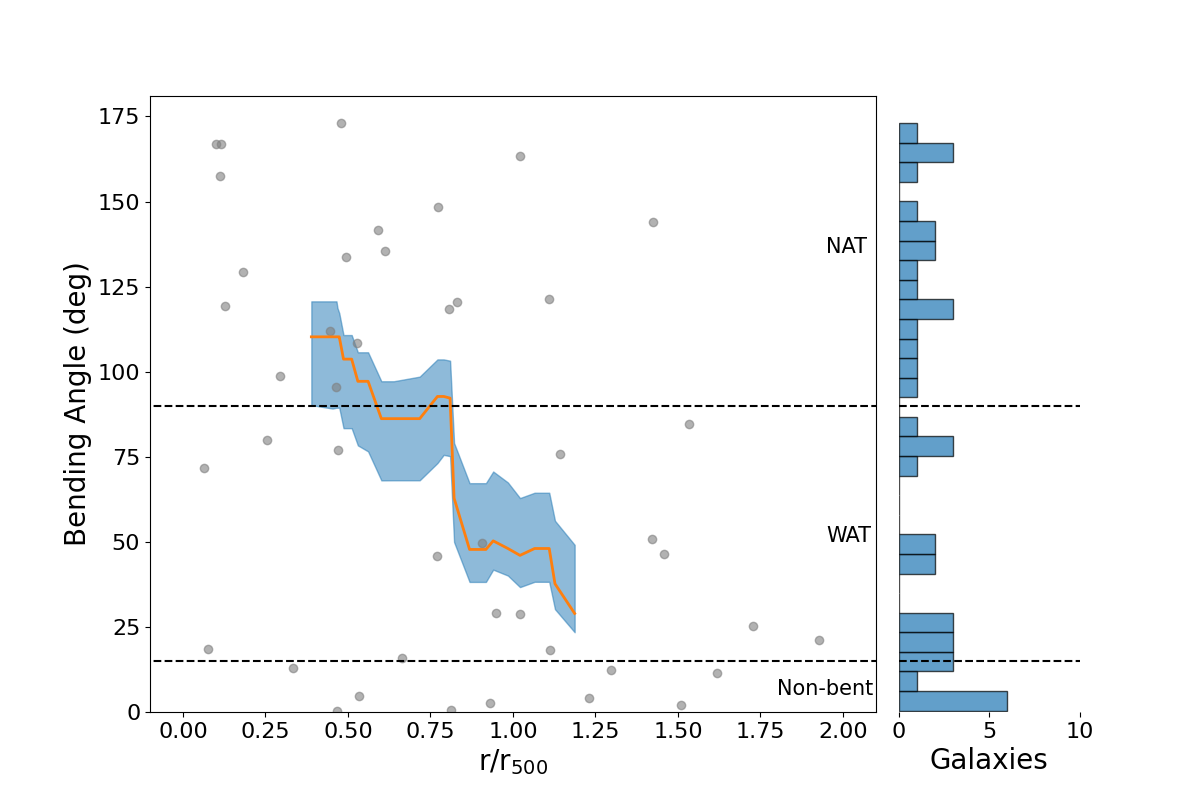}
  \caption{Tailed radio galaxy bending angle plotted against distance from the cluster centre, normalised by $r_{500}$. The grey markers show the individual tailed radio galaxies. Tailed radio galaxies above (below) the dashed line at \ang{90} are classified as NATs (WATs). The red line shows the running median. To calculate the running median, we followed \citet{2016ApJ...829....5L} and calculated the running median in bins of $N =$ 20 galaxies. The blue shaded area displays the 1$\sigma$ statistical uncertainty on the running median, calculated as $|M-[p16,p84]|/\sqrt{N}$, where $M$ is the running median and $p16$ and $p84$ are the 16 and 84 percentiles.}
     \label{fig:BAvsDistances}
\end{figure}

The fact that NATs are located closer to the cluster centres than WATs and the general population of galaxies suggests that galaxies closer to the cluster centre are more bent. In {\color{blue}\autoref{fig:BAvsDistances}}, we show the bending angle as a function of distance from the centre. Here, we only plot the radio galaxies for which both tails are resolved and the bending angle could be determined accurately. We find a decrease in median bending angle as a function of the distance in {\color{blue}\autoref{fig:BAvsDistances}}. This indicates that there is a negative correlation between the bending angle of tailed radio galaxies and their distance from the cluster centre. Correlation tests show a significant negative correlation, with Pearson $r=-0.41$ (p $=$ 0.006) and with Spearman $r=-0.36$ (p$=$0.02).

\begin{figure*}[tbh]
    \resizebox{\hsize}{!}
    {\includegraphics[width=\hsize]{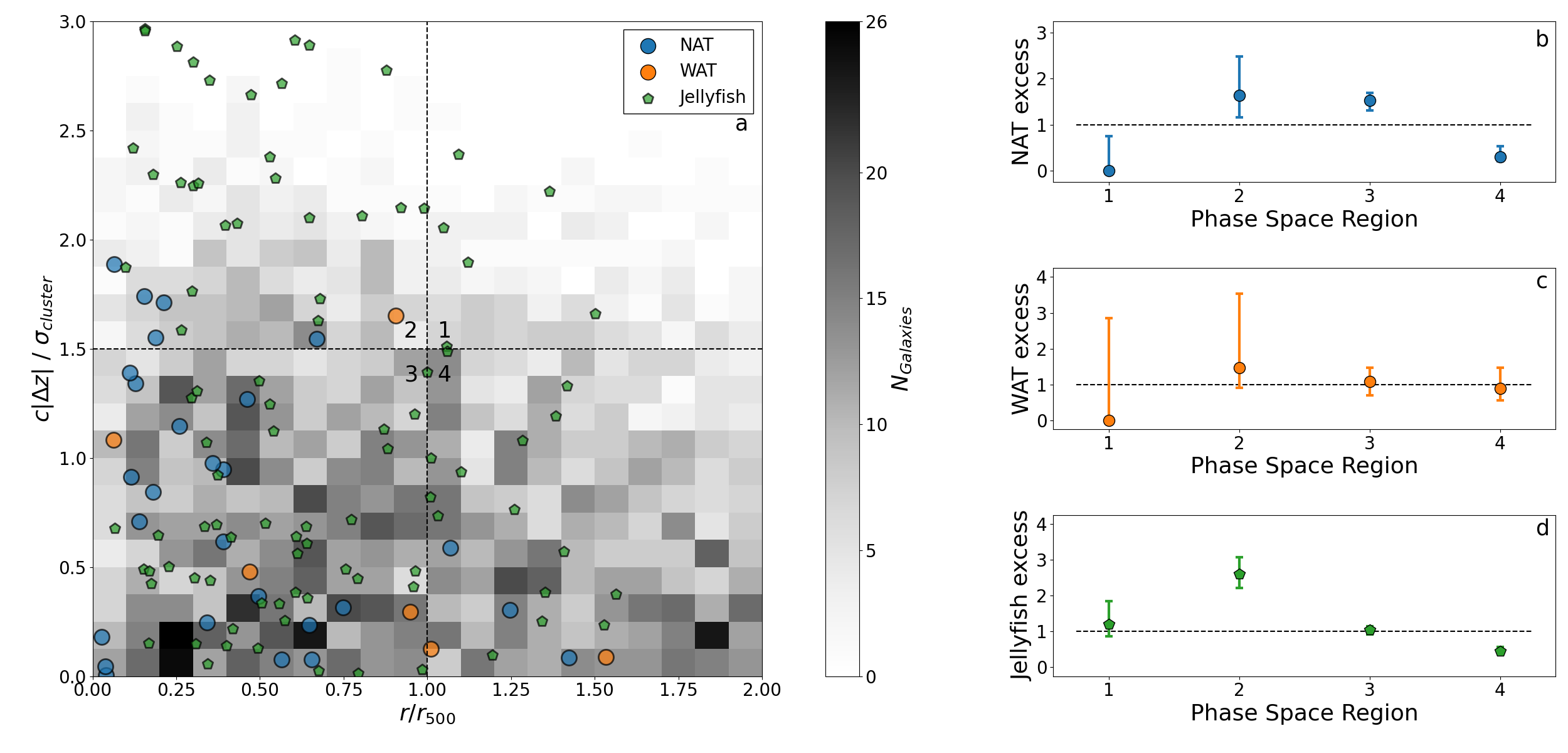}}
    \caption{Cluster phase-space diagram, showing excess for tailed radio galaxies. We show NATs and WATs in orange and blue, respectively. Panel (a): Phase-space diagram showing WATs, NATs, and jellyfish galaxies. In the background, the numbers of galaxies from HeCS-SZ \citep{2016ApJ...819...63R} are given as a reference. The dashed lines show the difference between the four different phase-space regions. Panel (b)-(d): Excess for NATs, WATs, and jellyfish galaxies as defined in \autoref{eq:excess}. The dashed line represents no excess. The error bars are the 1$\sigma$ statistical uncertainty following \citet{2021A&A...650A.111R}.}
    \label{fig:PhaseSpace}
\end{figure*}

\subsection{Phase space of tailed radio galaxies}\label{sec:Results_phase_space}
To study the orbital history of galaxies in the cluster, phase-space diagrams are commonly used \citep[e.g.][]{2017ApJ...843..128R,2021A&A...650A.111R}. We combined our results for the distances of tailed radio galaxies from the cluster centres with their velocities in a phase-space diagram in {\color{blue}\autoref{fig:PhaseSpace}a}. Additionally, we compared the tailed radio galaxies in our sample with the jellyfish galaxies from \citet{2021A&A...650A.111R}. Since \citet{2021A&A...650A.111R} report projected radii in terms of r/r$_\mathrm{180}$, we converted these radii to r/r$_\mathrm{500}$ using a conversion factor of 1.57, which assumes a Navarro-Frenk-White profile for the dark matter halo with a concentration parameter $c=5$ \citep{Pierpaoli2003}. We note that the sample from \citet{2021A&A...650A.111R} is cut off at r$_\mathrm{180}$. However, because the excess in a region of phase space is calculated as a ratio (cf. Eq \ref{eq:excess}), this simply lowers the sample size but does not bias the excess calculation. The main caveat of the phase-space analysis is that both the velocity and distance are lower limits of their real values. The physical intracluster velocities of the galaxies were taken to be their redshift offsets from that of their host cluster. We used spectroscopic redshifts from the HeCS-SZ \citep{2016ApJ...819...63R} and obtained physical velocities for galaxies in 24 of the 81 galaxy clusters in this study. 

We analysed the resultant phase-space diagram in {\color{blue}\autoref{fig:PhaseSpace}a}, adopting the method from \citet{2021A&A...650A.111R}, who analysed the phase-space diagram for jellyfish galaxies in four different quadrants of cluster phase space. The lines r/r$_\mathrm{500}=1$ and c$|\Delta$ z$|/\sigma_{cluster}=1.5$ divide the phase space in four quadrants, where c$|\Delta$ z$|/\sigma_{cluster}=1.5$ is the velocity offset to the cluster velocity normalised by the cluster dispersion from the HeCS-SZ \citep{2016ApJ...819...63R}. The number of NATs, WATs, and galaxies in the HeCS-SZ for each quadrant are shown in {\color{blue}\autoref{table:excess}}. From the number of galaxies in {\color{blue}\autoref{table:excess}} and \autoref{eq:excess}, the excess is calculated as
\begin{equation}\label{eq:excess}
    \eta = \left. {\left(\frac{N^{Q_{i}}_{Tails,DS}}{N_{Tails,DS}} \right)} \middle/ {\left( \frac{N^{Q_{i}}_{HeCS-SZ,DS}}{N_{HeCS-SZ,DS}} \right)} \right. ,
\end{equation}

where $\eta$ is the tailed radio galaxy excess and $N_{HeCS-SZ,DS}$ and $N_{Tails,DS}$ are the total numbers of galaxies selected from the HeCS-SZ sample and the total number of tailed radio galaxies for a given dynamical state $DS$ in the sample, respectively. $N^{Q_{i}}_{Tails,DS}$ and $N^{Q_{i}}_{HeCS-SZ,DS}$ are the total number of tailed radio galaxies and galaxies in the HeCS-SZ sample for each quadrant. A value of $\eta>1$ means that there is an excess and a value of $\eta<1$ means that there is a deficit in the number of tails in a phase-space region.

In {\color{blue}\autoref{fig:PhaseSpace}b-d} and {\color{blue}}\autoref{table:excess} we show the excess for NATs, WATs, and the jellyfish galaxies from \citet{2021A&A...650A.111R}. It is clear from the uncertainties in {\color{blue}\autoref{fig:PhaseSpace}c} that there are not enough WATs to draw significant conclusions. Thus, hereafterwe only discuss NATs. In phase-space region 1 (high velocity and large distance), we do not observe any NATs. Phase-space region 2 (high velocity and small distance), shows a hint of an excess in NATs ($\eta_{NAT,2}=$1.63$^{+0.85}_{-0.47}$). In phase-space region 3 (low velocity and small distance), the region with the most galaxies in the sample, we find a significant excess of NATs ($\eta_{NAT,3}=$1.53$^{+0.16}_{-0.22}$). Finally, phase-space region 4 (low velocity and large distance), shows a significant deficit in NATs $\eta_{NAT,4}=$0.30$^{+0.24}_{-0.09}$.

In summary, comparing the NATs with the general cluster population, we find that the most distinguishing phase-space property of NATs is that they live closer to the cluster centre than the other galaxies in the cluster. We find that the NATs (c$|\Delta$z$|$/$\sigma_{cluster}=$0.71$\pm$0.11) do not have velocities that differ significantly from the general cluster population (c$|\Delta$z$|$/$\sigma_{cluster}=$0.79$\pm$0.01). Overall we conclude that NATs are located in a different part of the phase space than the general cluster population, with similar velocities to the general population, but generally closer to the cluster centre, where the gas density and the effects of ram pressure are largest.

In {\color{blue}\autoref{fig:PhaseSpace}} and {\color{blue}}\autoref{table:excess} we compare the NATs with jellyfish galaxies. In phase-space region 1, no NATs are observed, {while the jellyfish galaxies are consistent with the cluster population ($\eta_{jellyfish,1}=$1.19$^{+0.66}_{-0.33}$). Phase-space region 2 shows similar results for NATs and jellyfish galaxies ($\eta_{jellyfish,2}=$2.60$^{+0.46}_{-0.40}$). In contrast, phase-space region 3 shows differences between the populations of NATs and jellyfish galaxies. NATs show an excess in this phase-space region, while jellyfish galaxies are consistent with the cluster population ($\eta_{jellyfish,3}=$1.03$^{+0.11}_{-0.11}$). In phase-space region 4, we find a deficit for both NATs and jellyfish galaxies ($\eta_{jellyfish,4}=$0.45$^{+0.12}_{-0.09}$). In summary, NATs live in a phase space that is more similar to jellyfish galaxies than to the general cluster population, but with the notable difference in phase-space regions 1 and 3.

\subsection{Orientation of tailed radio galaxies}
Tailed radio galaxies up to $7r_{500}$ have been found to be predominantly on radially inbound orbits \citep{2021MNRAS.506L..55D}. In this work, we determined the orientation of the tailed radio sources with respect to the cluster centre using the full sample of 100 NATs and WATs from {\color{blue}\autoref{table:cluster_member}}. We defined the orientation angle of the tailed radio source as the angle between the line from the cluster centre to the tailed radio galaxy host and the line from the galaxy host to the middle of both radio lobes. Thus, an orientation angle of \ang{0} is outbound and an angle of \ang{180} is inbound. 

In {\color{blue}\autoref{fig:orientation_hist}} we show the orientation angles for NATs and WATs. We find no significant preference for an orientation for both NATs and WATs. A KS-test between the orientation of tailed radio galaxies and a uniform distribution shows that there is no significant difference between the distributions of orientation for NATs (p $=$ 0.146) and WATs (p $=$ 0.076). This result does not support the findings of \citet{2021MNRAS.506L..55D}, which however considered a sample of tailed radio galaxies extending at larger distances than what is done here. Following \citet{2017ApJ...843..128R}, galaxies closer to the cluster centre could be past their first pericentre on a trajectory that is not a radially inbound orbit anymore. Hence, galaxies closer to the cluster centre have a less preferred orientation than galaxies further away from the cluster centre.

\begin{figure}[t]
\centering
\includegraphics[width=\hsize]{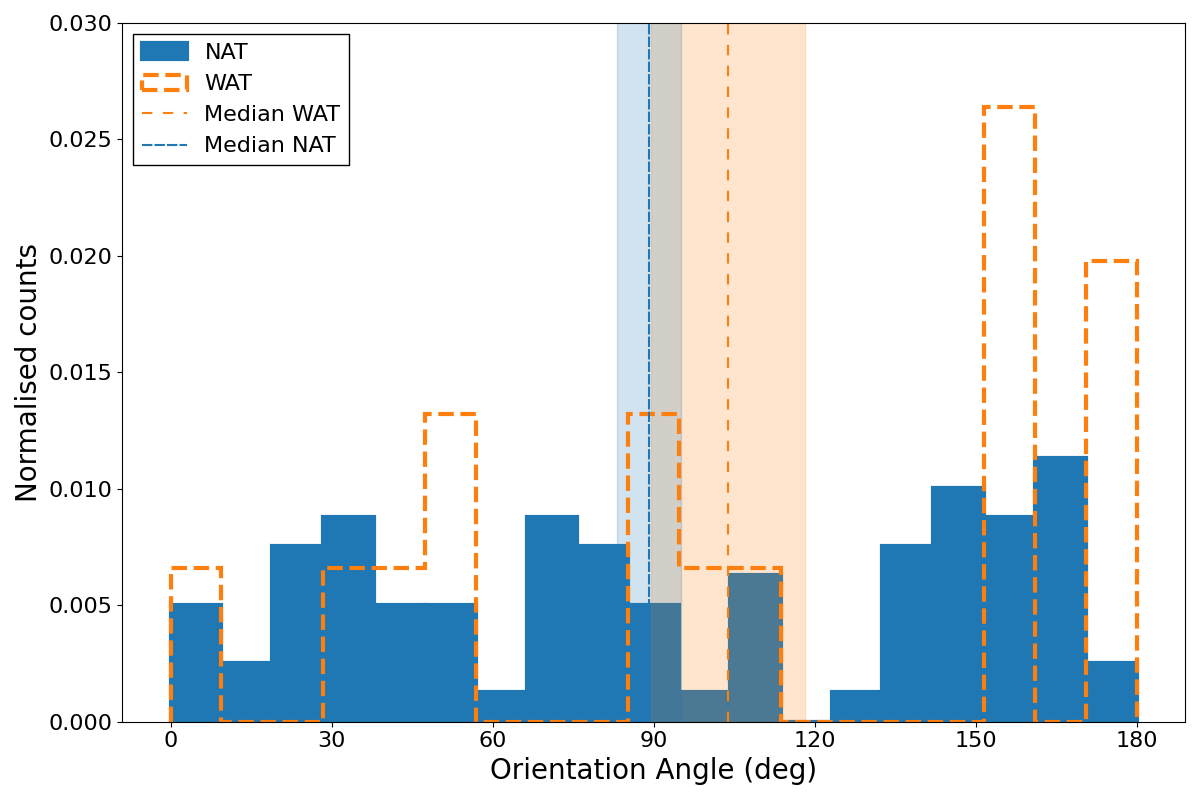}
  \caption{Histogram of the orientation angles for NATs and WATs. The orientation angle of \ang{0} is outbound and \ang{180} is inbound. The dashed lines show the median values of the orientation angles, and the shaded region gives the 1$\sigma$ statistical uncertainty on the median. The counts for each histogram are normalised such that the area under each histogram integrates to 1.}
     \label{fig:orientation_hist}
\end{figure}

\subsection{Dynamical state}
It is been suggested that merging galaxy clusters host more tailed radio galaxies than relaxed clusters, as merging clusters have larger systematic gas motions \citep{1999MNRAS.305..417S,Morris_2022}. From X-ray data, the dynamical states of 62 clusters in this study are known \citep{2017ApJ...843...76A,Morris_2022}. To determine if a galaxy cluster is merging or relaxed we used the concentration parameter between $0.15r_{500}$ and $r_{500}$ ($C_{SB}$) and the break value of 0.4 from \citet{2017ApJ...843...76A}. The concentration parameter is

\begin{equation}
    C_{SB} = \frac{\sum(<0.15r_{500})}{\sum(<r_{500})} ,
\end{equation}

\noindent where $\sum(<r)$ is the integrated projected X-ray emissivity within a circle of radius $r$ \citep{2017ApJ...843...76A}.

In {\color{blue}\autoref{tab:dynamicalstates}} we list the average number of WATs and NATs for relaxed and merging galaxy clusters. We find no significant differences in the total number of WATs and NATs between the full sample of 81 clusters, the sample of 19 relaxed clusters, and the sample of 44 merging clusters. This is unlike the findings of \citet{Bliton:1998mi}, where they state that NATs are located in dynamically complex systems by preference. Comparably, the fact that the average number of WATs is similar in relaxed and merging clusters is not in line with the findings from \citet{1994AJ....108.2031P}, \citet{1995ApJ...445...80L}, \citet{Gomez1997a}, \citet{Gomez1997b}, and \citet{2000MNRAS.311..649S} who suggest that WATs are a by-product of merging galaxy clusters. We find that the results do not significantly differ when using other dynamical parameters than $C_{SB}$, as defined in \citet{2017ApJ...843...76A}, and conclude that we find no significant evidence for WATs or NATs to occur preferentially in clusters with specific dynamical states.

We display the excess of galaxies in phase space for merging and relaxed clusters in {\color{blue}\autoref{fig:PhaseSpaceDynState}} and {\color{blue}\autoref{table:excessDynState}}. For WATs we cannot make decisive conclusions because of the low number of WATs in this sample. However, we find an excess of NATs in phase-space region 2 (high velocity and small distance) for NATs in merging clusters ($\eta_{NAT,merging,2}=$2.19$^{+1.17}_{-0.69}$), while there is a hint of a deficit for NATs in relaxed clusters ($\eta_{NAT,relaxed,2}=$0.00$^{+0.87}_{-0.00}$). In phase-space region 3 (low velocity and small distance), we find an excess of NATs in relaxed clusters ($\eta_{NAT,relaxed,3}=$2.22$^{+0.08}_{-0.35}$), while the excess of NATs in merging clusters is consistent with the general cluster population in merging clusters ($\eta_{NAT,merging,3}=$1.18$^{+0.23}_{-0.26}$). In phase-space region 4 (low velocity and large distance), we find a deficit for NATs in both merging ($\eta_{NAT,merging,4}=$0.51$^{+0.36}_{-0.17}$) and relaxed ($\eta_{NAT,relaxed,4}=$0.00$^{+0.27}_{-0.00}$) clusters. In summary, the phase-space region showing the strongest excess in NATs seems to be dependent on the merging state of the clusters. Although the average number of NATs and WATs is consistent between merging and relaxed clusters, merging clusters show the strongest excess of NATs in phase-space region 3, while relaxed clusters show the strongest excess in phase-space region 2.

\begin{table}[b]
    \centering
    \caption{Average number of WATs or NATs for a galaxy cluster.}
    \label{tab:dynamicalstates}
    \begin{tabular}{c c c}
    \hline
                                        & WAT                           & NAT                           \\ \hline
    Full Sample (81 clusters)  & 0.20 ± 0.05 & 1.04 ± 0.11 \\
    Relaxed (19 clusters)      & 0.11 ± 0.07 & 1.26 ± 0.26 \\
    Merging (44 clusters)      & 0.20 ± 0.07 & 1.18 ± 0.16 \\ \hline
    \end{tabular}
\end{table}

\section{Discussion}\label{sec:Discussion}
\subsection{Phase space of tailed radio galaxies}
Earlier work suggests that extended radio sources evolve as they fall into the cluster \citep[e.g.][]{1972Natur.237..269M,1976ApJ...205L...1O}. We observe that tailed radio galaxies with large bending angles generally live closer towards the cluster centre than tailed radio galaxies with small bending angles, as shown in {\color{blue}\autoref{fig:Distances}}. Additionally, we find a negative correlation between the bending angle and the distance from the cluster centre, as presented in {\color{blue}\autoref{fig:BAvsDistances}}. Our results are in line with results from previous studies \citep[e.g.][]{2019AJ....157..126G, 2021MNRAS.506L..55D}, which show that the tails are more bent when they are located closer to the  cluster centre.

Ram pressure is related to the density of the cluster gas and the velocity of the radio source following
\begin{equation}\label{eq:ram_press}
    P_{ram} \propto \rho_{gas} v^2,
\end{equation}
where $P_{ram}$ is the ram pressure, $\rho_{gas}$ is the density of the ICM and $v$ is the relative velocity of the host galaxy \citep{1979Natur.279..770B,1979ApJ...234..818J}. The density of the cluster gas is dependent on the distance from the cluster centre. Gas closer to the cluster centre is denser. Therefore, radio galaxies closer to the cluster centre generally live in denser environments. The results in this work support that the bending angle of tailed radio galaxies is linked to the ram pressure that is exerted on the galaxy as it moves through the ICM \citep{1979Natur.279..770B,1979ApJ...234..818J}.

Previous studies were able to show the relation between the bending angle and the distance from the cluster centre for larger samples \citep{2019AJ....157..126G,2021MNRAS.506L..55D}. However, the relation between the bending angle and the velocity offset from the cluster has only been investigated for smaller samples. \citet{1972Natur.237..269M} investigated tailed radio galaxies in the Coma, Perseus, and the 3C129 clusters. They suggested that the morphology of some of the tailed radio galaxies in these clusters via ram pressure stripping can be explained due to their high radial velocity offsets with respect to the cluster. \citet{1978ApJ...221..422U} found similar results for 4 clusters of galaxies. In this work, we establish a better view of the relation between the bending of radio galaxies and ram pressure by showing the phase space of tailed radio galaxies in {\color{blue}\autoref{fig:PhaseSpace}}.

Due to low number counts, the phase-sapce results for WATs are inconclusive. We find an excess in phase-space regions 2 (high velocity and small distance) and 3 (low velocity and small distance) for NATs. No NATs were detected in region 1 (high velocity and large distance). Following \autoref{eq:ram_press}, the regions of phase space from most to least ram pressure in {\color{blue}\autoref{fig:PhaseSpace}} are phase-space region 2, 3, 1, and 4. This suggests that the highly bent NATs live in dense environments with the most ram pressure and do not necessarily have higher velocities than other cluster members.

Comparing the results of NATs in phase space with those of the general cluster population clearly shows the influence of ram pressure on the bending angle of tailed radio galaxies. Figure 1 in \citet{2017ApJ...843..128R} shows a typical trajectory of a galaxy that is falling into a galaxy cluster. Following this typical trajectory, we suggest a plausible scenario for the evolution of the bending angle of a tailed radio galaxy. This evolutionary trajectory of tailed radio galaxies within galaxy clusters can be outlined as follows: Initially, the radio galaxy starts as a non-bent double lobe or FRI plume in phase-space region 4 (low velocity and large distance). As the galaxy falls into the gravitational well of the cluster, the galaxy moves more to phase-space regions 2 and 3, and the lobes gradually bend, forming a NAT morphology. This bending process intensifies as the galaxy approaches closer to the cluster centre, leading to the formation of a NAT in phase-space region 2. Subsequently, as the galaxy settles in the cluster in phase-space region 3, both its velocity and bending angle diminish. The excess of NATs in the high velocity and small distance region found in {\color{blue}\autoref{fig:PhaseSpace}} is in line with the expectations from earlier studies \citep[e.g.][]{1972Natur.237..269M}. Combined with the negative correlation we find for bending angle and distance ({\color{blue}\autoref{fig:PhaseSpace}}), this supports the idea that the bending angle is linked to the ram pressure.

\subsection{Comparison with jellyfish galaxies}
Jellyfish galaxies are another phenomenon where galaxies in clusters are affected by ram pressure \citep{1972ApJ...176....1G,2000ApJ...540..113B,2014MNRAS.445.4335F,2016AJ....151...78P}. The ram pressure strips the ISM as the galaxy travels through the ICM. This leads to gas tails. There is evidence of these observed gas tails from radio regimes to X-ray \citep[e.g.][]{2019MNRAS.482.4466P}. We compared, in {\color{blue}\autoref{sec:Results_phase_space}}, our results to a sample of jellyfish galaxies from lower-redshift clusters ($z<0.05$) from \citet{2021A&A...650A.111R}.

We find, that jellyfish galaxies and tails are located in similar parts of the phase space, but with the notable difference that tails show an excess close to the cluster centre (phase-space region 3) while jellyfish galaxies show an excess away from the cluster centre (phase-space region 1) where no tailed radio galaxies were found. The absence of NATs and WATs in phase-space region 1, characterised by high velocity and large distance from the cluster centre, could be because the density at these large distances is not high enough to significantly cause bending. Although this is not necessarily supported by \citet{2021MNRAS.506L..55D} who found tails out to large distances, a crucial difference is that this study was limited to radii up to 2$r_\mathrm{500}$. It seems likely that the absence of tailed radio galaxies is simply due to the relatively small sample size in this region. Considering that the total number of galaxies inhabiting this region is already low, i.e. only $\sim4.6\%$ of the cluster galaxies live in this region, the uncertainty is too large. Over all regions of phase space, NATs and WATs respectively constitute only $\sim0.6\%$ and $\sim0.3\%$ of the total galaxy population. Conservatively applying this scaling to region 1, we would expect at most two NATs and one WAT in phase-space region 1. Additionally, the radio luminosity of the lobes is correlated with the density of the environment surrounding the galaxy \citep{2013MNRAS.430..174H}. This also makes (tailed) radio galaxies in phase-space region 1 intrinsically more difficult to observe. As these numbers of NATs and WATs are small, we conclude that a larger sample of tailed radio galaxies is needed to make definite conclusions about the absence of tailed radio galaxies in phase-space region 1.

In contrast, the non-detection of an excess of jellyfish galaxies in phase-space region 3, where a significant excess of tailed radio galaxies can be found can be explained by the difference between the two classes of sources. The jellyfish galaxy tail originates from ram pressure stripping of the ISM in star-forming galaxies. As jellyfish galaxies fall inside the cluster, the available material depletes, and as the galaxies settle in the cluster most of the gas is stripped off already. Jellyfish galaxies tend to be skewed to larger velocity offsets, this is consistent with them being on their first infall towards the cluster centre \citep{2022ApJ...934...86S}. The star formation of galaxies in clusters is quenched relatively quickly and therefore jellyfish galaxies do not survive much beyond their first pericentric passage \citep{2021A&A...650A.111R,2021A&A...652A.153R}. On the contrary, in radio galaxies, the tail originates from the AGN, which undergoes a series of active cycles that can last for $\sim 10^{7}$ yr \citep[e.g.][]{2017NatAs...1..596M}. Tailed radio galaxies lose their material with time as well because the AGNs is fuelled by the galaxy gas reservoir. Moreover, tailed radio galaxies lose their gas reservoirs due to ram pressure, but while there is evidence that the fraction of AGN decreases as they fall towards the centre of clusters \citep[e.g.][]{2013MNRAS.429.1827P,2020AJ....159...69M,2024arXiv240105747K}, it is clear from our results that tailed radio galaxies can sustain a supply of material that is being ejected from the host galaxy, while in jellyfish galaxies this supply is depleted more quickly. \citet{2021A&A...650A.111R,2021A&A...652A.153R} report that most of the jellyfish tails are directed in the opposite direction to the centre, thus suggesting that the jellyfish galaxies get stripped during the first infall. In {\color{blue}\autoref{fig:orientation_hist}}, we show that tailed radio galaxies do not have a preferential orientation, suggesting that tailed radio galaxies can survive the first infall. Therefore, it is possible to have a radio galaxy that has been a long-time cluster member but that still shows a tail. This is consistent with the radio galaxies being more concentrated towards the core of the phase space.

\subsection{Cluster properties}\label{sec:disc_cluster_prop}
Earlier studies suggested that tailed radio galaxies might form due to the merging of galaxy clusters \citep[e.g.][]{Bliton:1998mi,2000MNRAS.311..649S}. However, {\color{blue}\autoref{tab:dynamicalstates}} shows that there is no significant difference in the average number of WATs and NATs in merged and relaxed clusters. This result is consistent with the findings in \citet{2023MNRAS.526.4831L}. \citet{2023MNRAS.526.4831L} studied galaxies undergoing ram pressure stripping in 52 galaxy clusters. They found no clear correlation between ram pressure stripping and the cluster dynamical state. This supports that both tailed radio galaxies and jellyfish galaxies are defined by ram pressure.

Interestingly, {\color{blue}\autoref{fig:PhaseSpaceDynState}} also shows that the excess of NATs in phase-space region 2 (high velocity and small distance) can be mainly attributed to an excess in merging clusters, while the excess of NATs in phase-space region 3 (low velocity and distance) is mainly caused by the sample of relaxed clusters. Furthermore, we do not find any NATs or WATs from relaxed clusters outside of phase-space region 3. Based on the findings in {\color{blue}\autoref{fig:PhaseSpaceDynState}}, we cautiously propose the following hypothesis. After falling into the cluster, the tailed radio galaxies settle according to the track shown in \citet[][Fig. 1]{2017ApJ...843..128R} towards the centre of the cluster (phase-space region 3). For relaxed clusters, the tailed radio galaxies remain there, but in merging galaxy clusters the galaxies get stirred up by the merger. This pushes them towards higher velocity regions of the phase space with respect to the cluster centre (phase-space region 2). The radio galaxies will eventually follow a similar track again as the cluster relaxes, and they travel to the lower-velocity and large-distance regions (phase-space region 4), back-splashing, and eventually settling again in the cluster centre. However, this would also mean that the other galaxies in the cluster should follow a similar path. Therefore, the excess of NATs should not change for merging clusters unless there is some mechanism preferentially stirring up galaxies with bent radio jets. This would result in different orientations of NATs in merging and relaxed clusters. However, we checked the orientation angle by splitting on cluster dynamical state and did not observe a significant difference between the orientation of NATs in merging and relaxed clusters. It is possible that a larger sample of tailed radio galaxies is required to confirm this theory.

We note that X-ray clusters samples are biased towards relaxed clusters \citep[e.g.][]{1990ASSL..160..372P,2011A&A...526A..79E}, while the SZ-selected sample from \textit{Planck} used in this work are more representative of the distribution of the dynamical states of the general cluster population \citep{2017ApJ...843...76A,2017MNRAS.468.1917R}. However, there are more factors that could lead to biases in the results in {\color{blue}\autoref{tab:dynamicalstates}} than cluster dynamical state alone. To check whether the host clusters have similar properties for merging and relaxed clusters, we plotted the redshift and mass distributions for the merging and relaxed galaxy clusters (see Fig. \ref{fig:distrb_clusters}). The median redshift for both relaxed ($z_{relaxed} = 0.206$±$0.015$) and merging clusters ($z_{merging} = 0.210$±$0.011$) is not significantly different. The median mass for merging clusters ($M_{merging} = $ ($6.300$±$0.280$)$\times10^{14}M_\sun$) is lower than for relaxed clusters ($M_{relaxed} = $ ($7.578$±$0.264$)$\times10^{14}M_\sun$). However, based on a two-sample KS-test on mass and redshift between merging and relaxed clusters we find that there is no significant difference between the population of merging and relaxed clusters in mass (p $=$ 0.22) and redshift (p $=$ 0.39).
Therefore, we conclude that, although the occurrence of WATs and NATs does not seem to depend on dynamical state, the phase-space distribution of WATs and NATs might be linked to the dynamical state of the host clusters, although a larger sample is needed to confirm whether the cluster mass is not a confounding variable. This hints at a possibly interesting effect between the bending of tails and the host cluster dynamical state.

In this work, we started from a well-defined and mass-limited sample of X-ray-detected \textit{Planck} clusters at low redshifts ($z<0.35$). These clusters are relatively massive ($M_\mathrm{500} > 4\times10^{14} M_\odot$, see Fig. \ref{fig:distrb_clusters}) and might thus not be representative of the typical environment of tailed radio galaxies. A recent study by \citet{Lao2025} identified 4876 tailed radio galaxies in the FIRST survey, of which 3087 were closely associated with clusters. However, they found that the mean mass of the host clusters is $1.5\times10^{14}M_\odot$, with only 6\% of tailed radio galaxies found in clusters above a mass of $3.2 \times 10^{14}M_\odot$. Additionally, a lower frequency of jellyfish galaxies and weaker ram pressure stripping of these galaxies are reported in lower-mass groups of galaxies \citep{2021A&A...652A.153R}. 

Finally, we emphasise that this study was limited to radio galaxies relatively close to the cluster centre ($r/r_\mathrm{500}<2.0$), while there is evidence that tailed radio galaxies are also influenced by clusters at much larger distances \citep[e.g.][]{2021MNRAS.506L..55D}.

\subsection{Future work}
This work focused on the effect of the cluster dynamical state and phase space of host galaxies on the bending angle of tailed radio sources. While we find that the position in phase space is most important for the amount of radio source bending, in some cases, the ICM magnetic field also has a large impact on the bending of radio jets \citep[e.g.][]{2021Natur.593...47C}. Numerical simulations show that the magnetic fields also have an impact on the bending of the radio jets \citep{2017ApJ...839...14G,2019A&A...621A.132M,2022A&A...659A.139M}. It would be interesting to investigate to what extent the cluster or host galaxy magnetic fields play a role in the bending of tailed radio galaxies \citep{2013MNRAS.432..243P,2021MNRAS.508.5326M}. This will be investigated in a future paper, as polarisation information is available for these sources \citep{2022A&A...665A..71O}.

Another extension of this study would be to probe lower-mass clusters and the presence of NATs and WATs at higher redshifts. This could provide important information on the build-up and evolution of the ICM. In addition, combining all-sky galaxy cluster and group catalogues from X-ray surveys such as eROSITA \citep{2024arXiv240208452B} with large area radio surveys such as LOFAR \citep{2017A&A...598A.104S,2019A&A...622A...1S,2022A&A...659A...1S}, VLASS \citep{Lacy_2020}, or EMU \citep{2011PASA...28..215N,2021PASA...38...46N} can provide orders of magnitude larger samples in the near future.

\section{Conclusions}\label{sec:Conclusion}
In this work we identified 127 extended radio sources from VLA observations of 81 galaxy clusters. Among these, we found 84 NATs and 16 WATs, and 9 non-bent radio galaxies, classifying the sources based on their bending angle. For a subset of sources present in the HECS-SZ catalogue, we calculated the velocity offsets from the cluster centre, and the dynamical states were available for 62 clusters using \textit{Chandra} observations from \citet{2017ApJ...843...76A}. The conclusions of this work are: 

\begin{enumerate}
    \item NATs are located closer the cluster core than the regular galaxies and WATs within a galaxy cluster, with a median radius >5$\sigma$ below WATs and the regular galaxies. This is in line with results from studies that selected tails and subsequently classified their environment \citep[e.g.][]{2019MNRAS.488.2701M,2019AJ....157..126G}.
    \item There is a significant negative correlation between the bending angle and the distance from the cluster centre, i.e. radio galaxies are systematically bent more extremely closer to the centre of the cluster. Pearson and Spearman correlation tests give $r=-0.41$ (p $=$ 0.006) and $r=-0.36$ (p$=$0.02), respectively.
    \item NATs show a slight excess in the high-velocity and low-distance region and a significant excess in the low-velocity and low-distance phase-space region ($\eta_{NAT,2}=$1.63$^{+0.85}_{-0.47}$ and $\eta_{NAT,3}=$1.53$^{+0.16}_{-0.22}$).
    \item NATs behave more similarly in phase space to jellyfish galaxies than the general cluster population. We find a (slight) excess of NATs in phase-space regions 2 (high velocity and small distance) and 3 (low velocity and small distance), while \citet{2021A&A...650A.111R} find an excess of jellyfish galaxies only in phase-space region 2  ($\eta_{jellyfish,2}=$2.60$^{+0.46}_{-0.40}$). This suggests that ram pressure plays a key role in both the stripping of gas and the bending of radio tails. Jellyfish galaxies being quenched on their first infall into the cluster could be an explanation for the lack of excess in phase-space region 3, an effect that does not impact NATs as strongly.
    \item WATs and NATs have the same occurrence ratios in merging and relaxed clusters. However, they are distributed differently in phase space for clusters of different merger states. In merging clusters NATs have the largest excess in phase-space region 2 ($\eta_{NAT,merging,2}=$2.19$^{+1.17}_{-0.69}$), while for relaxed clusters this is phase-space region 3 $\eta_{NAT,relaxed,3}=$2.22$^{+0.08}_{-0.35}$. We do not observe any tailed radio galaxies in relaxed clusters outside of phase-space region 3, while they are found in other phase-space regions in merging clusters.

\end{enumerate}

\bibliographystyle{aa}
\bibliography{main.bib}

\section*{Data availability}
\autoref{tab:A1}  only available in electronic form at the CDS via anonymous ftp to \url{cdsarc.u-strasbg.fr (130.79.128.5)} or via \url{http://cdsweb.u-strasbg.fr/cgi-bin/qcat?J/A+A/}.

\begin{acknowledgements}

We would like  to  thank  the  anonymous referee for useful comments. 
We thank L. Rudnick for his helpful comments on this work

RJvW and EO acknowledge support from the VIDI research programme with project number 639.042.729, which is financed by the Netherlands Organisation for Scientific Research (NWO). IDR acknowledges support from the Banting Fellowship Program. AB acknowledges financial support from the European Union - Next Generation EU. AI aknowledges funding from the European Research Council (ERC) under the European Union's Horizon 2020 research and innovation programme (grant agreement No. 833824, PI Poggianti) and the INAF founding program 'Ricerca Fondamentale 2022' (PI A. Ignesti).

LOFAR is the Low Frequency Array designed and constructed by ASTRON. It has observing, data processing, and data storage facilities in several countries, which are owned by various parties (each with their own funding sources), and which are collectively operated by the ILT foundation under a joint scientific policy. The ILT resources have benefited from the following recent major funding sources: CNRS-INSU, Observatoire de Paris and Université d’Orléans, France; BMBF, MIWF-NRW, MPG, Germany; Science Foundation Ireland (SFI), Department of Business, Enterprise and Innovation (DBEI), Ireland; NWO, The Netherlands; The Science and Technology Facilities Council, UK; Ministry of Science and Higher Education, Poland; The Istituto Nazionale di Astrofisica (INAF), Italy. This research made use of the Dutch national e-infrastructure with support of the SURF Cooperative (e-infra 180169) and NWO (grant 2019.056). The Jülich LOFAR Long Term Archive and the German LOFAR network are both coordinated and operated by the Jülich Supercomputing Centre (JSC), and computing resources on the supercomputer JUWELS at JSC were provided by the Gauss Centre for Supercomputing e.V. (grant CHTB00) through the John von Neumann Institute for Computing (NIC). This research made use of the University of Hertfordshire high-performance computing facility and the LOFAR-UK computing facility located at the University of Hertfordshire and supported by STFC [ST/P000096/1], and of the Italian LOFAR IT computing infrastructure supported and operated by INAF, and by the Physics Department of Turin university (under an agreement with Consorzio Interuniversitario per la Fisica Spaziale) at the C3S Supercomputing Centre, Italy. The data are published via the SURF Data Repository service which is supported by the EU funded DICE project (H2020-INFRAEOSC-2018-2020 under Grant Agreement no. 101017207)

The Pan-STARRS1 Surveys (PS1) and the PS1 public science archive have been made possible through contributions by the Institute for Astronomy, the University of Hawaii, the Pan-STARRS Project Office, the Max-Planck Society and its participating institutes, the Max Planck Institute for Astronomy, Heidelberg and the Max Planck Institute for Extraterrestrial Physics, Garching, The Johns Hopkins University, Durham University, the University of Edinburgh, the Queen's University Belfast, the Harvard-Smithsonian Center for Astrophysics, the Las Cumbres Observatory Global Telescope Network Incorporated, the National Central University of Taiwan, the Space Telescope Science Institute, the National Aeronautics and Space Administration under Grant No. NNX08AR22G issued through the Planetary Science Division of the NASA Science Mission Directorate, the National Science Foundation Grant No. AST-1238877, the University of Maryland, Eotvos Lorand University (ELTE), the Los Alamos National Laboratory, and the Gordon and Betty Moore Foundation. 

The Legacy Surveys consist of three individual and complementary projects: the Dark Energy Camera Legacy Survey (DECaLS; Proposal ID 2014B-0404; PIs: David Schlegel and Arjun Dey), the Beijing-Arizona Sky Survey (BASS; NOAO Prop. ID 2015A-0801; PIs: Zhou Xu and Xiaohui Fan), and the Mayall z-band Legacy Survey (MzLS; Prop. ID 2016A-0453; PI: Arjun Dey). DECaLS, BASS and MzLS together include data obtained, respectively, at the Blanco telescope, Cerro Tololo Inter-American Observatory, NSF’s NOIRLab; the Bok telescope, Steward Observatory, University of Arizona; and the Mayall telescope, Kitt Peak National Observatory, NOIRLab. The Legacy Surveys project is honoured to be permitted to conduct astronomical research on Iolkam Du’ag (Kitt Peak), a mountain with particular significance to the Tohono O’odham Nation. 

SDSS-IV is managed by the Astrophysical Research Consortium for the Participating Institutions of the SDSS Collaboration including the Brazilian Participation Group, the Carnegie Institution for Science, Carnegie Mellon University, Center for Astrophysics | Harvard \& Smithsonian, the Chilean Participation Group, the French Participation Group, Instituto de Astrof\'isica de Canarias, The Johns Hopkins University, Kavli Institute for the Physics and Mathematics of the Universe (IPMU) / University of Tokyo, the Korean Participation Group, Lawrence Berkeley National Laboratory, Leibniz Institut f\"ur Astrophysik Potsdam (AIP),  Max-Planck-Institut f\"ur Astronomie (MPIA Heidelberg), Max-Planck-Institut f\"ur Astrophysik (MPA Garching), Max-Planck-Institut f\"ur Extraterrestrische Physik (MPE), National Astronomical Observatories of China, New Mexico State University, New York University, University of Notre Dame, Observat\'ario Nacional / MCTI, The Ohio State University, Pennsylvania State University, Shanghai Astronomical Observatory, United Kingdom Participation Group, Universidad Nacional Aut\'onoma de M\'exico, University of Arizona, University of Colorado Boulder, University of Oxford, University of Portsmouth, University of Utah, University of Virginia, University of Washington, University of Wisconsin, Vanderbilt University, and Yale University.

This research has made use of the NASA/IPAC Extragalactic Database (NED), which is operated by the Jet Propulsion Laboratory, California Institute of Technology, under contract with the National Aeronautics and Space Administration.

This research has made use of NASA's Astrophysics Data System (ADS).       

This research has made use of data obtained from the \textit{Chandra} Data Archive and the \textit{Chandra} Source Catalog, and software provided by the \textit{Chandra} X-ray Center (CXC) in the application packages CIAO, ChIPS, and Sherpa. Based on observations obtained with XMM-Newton, an ESA science mission with instruments and contributions directly funded by ESA Member States and NASA.
\end{acknowledgements}
\onecolumn
\begin{appendix}
\section{Full source catalogue}

\begin{table}[h]
\centering
\caption{First 10 rows of the catalogue of the 127 extend radio sources that were detected in this work.}
\label{tab:A1}
\begin{tabular}{lllllllll}
\hline
RA [deg] & Dec. [deg] & z    & BA [deg] & r/r$_{500}$        & $\theta$[deg] & PLCKESZ       & BCG & c$|\Delta$z$|$/$\sigma_v$ \\
\hline \\ [-1.5ex]
357.9846 & -26.064 & 0.23   & 135.4  & 0.61      & 162.45      & G034.03-76.59 & no  & -        \\
354.4175 & 0.278   & 0.27   & 180.0  & 0.21      & 73.68       & G087.03-57.37 & no  & 1.71     \\
336.5093 & 17.3797 & 0.1    & 180.0  & 0.03      & 66.46       & G080.38-33.20 & no  & 0.18     \\
336.5046 & 17.3943 & 0.11   & 119.17 & 0.13      & 21.86       & G080.38-33.20 & no  & 1.34     \\
330.214  & 21.112  & 0.75   & 180.0  & 1.07      & 140.15      & G077.90-26.64 & no  & 0.59     \\
328.4035 & 17.6955 & 0.25   & -      & 0.0       & -           & G073.96-27.82 & yes & -        \\
323.8578 & -0.9625 & 0.32   & 46.3   & 1.46      & 176.9       & G053.44-36.26 & no  & -        \\
323.8503 & -0.9974 & 0.33   & -      & 1.02      & -           & G053.44-36.26 & no  & -        \\
323.8282 & 1.4242  & 0.23   & 18.42  & 0.08      & 55.36       & G055.97-34.88 & yes & -        \\
323.8143 & 1.4122  & 0.27   & 180.0  & 0.25      & 69.16       & G055.97-34.88 & no  & -        \\
\hline
\end{tabular}\\
\tablefoot{The full table is available in electronic form at the CDS. BA and $\theta$ give respectively the bending angle and the orientation angle to the cluster centre of the extended radio source. r/r$_{500}$ is the projected radius to the cluster centre and c$|\Delta$z$|$/$\sigma_v$ gives the galaxy velocity normalised by the cluster galaxy velocity dispersion.} 
\end{table}
\newpage
\section{Results by dynamical state}
\begin{figure*}[tbh]
    \resizebox{\hsize}{!}
    {\includegraphics[width=\hsize]{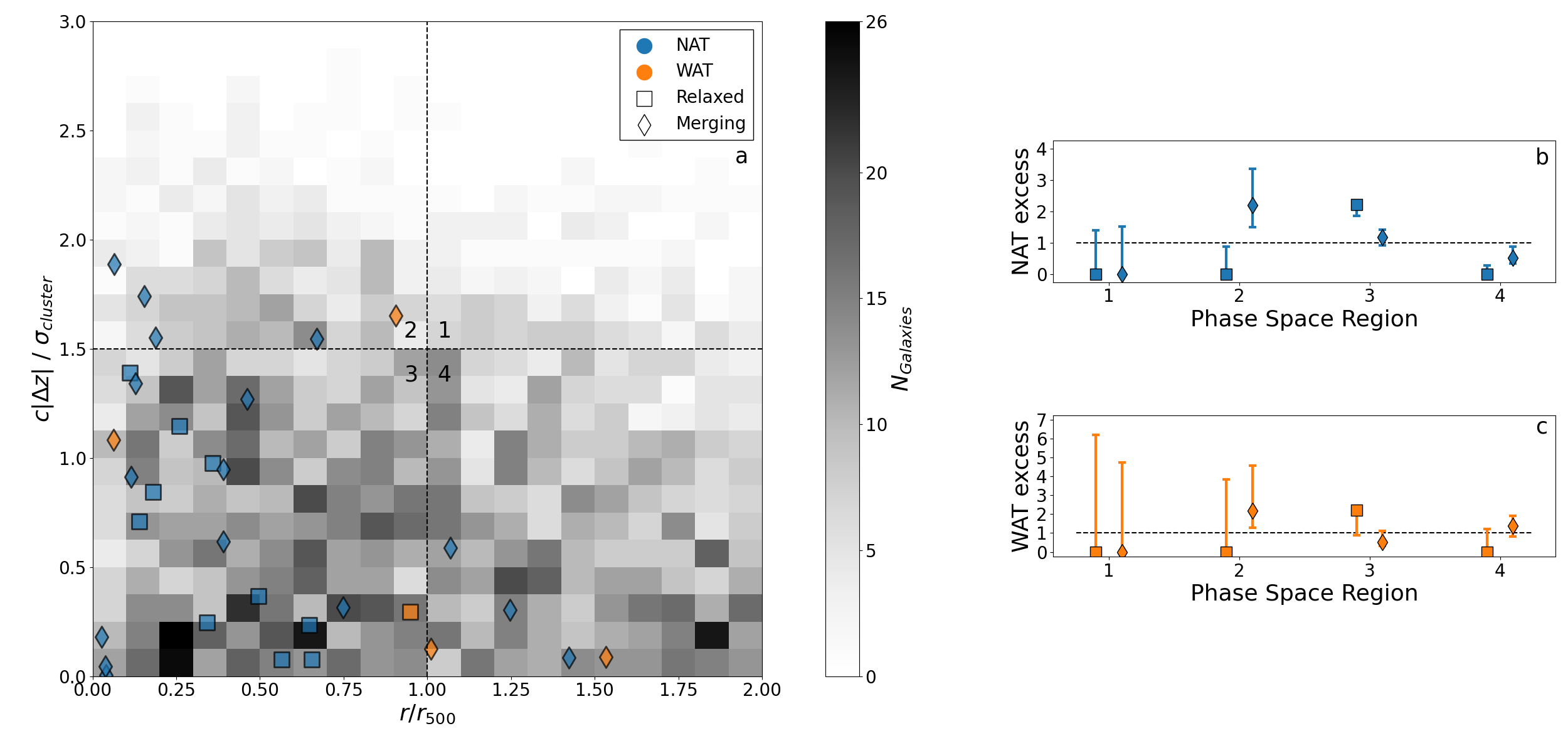}}
    \caption{Cluster phase-space diagram, showing excess for tailed radio galaxies in merging and relaxed clusters. We show NATs and WATs in orange and blue markers, respectively, and tailed radio galaxies in relaxed and merging clusters with respectively diamonds and squares. Panel (a): Phase-space diagram showing WATs and NATs. In the background, the numbers of galaxies from HeCS-SZ \citep{2016ApJ...819...63R} are given as a reference. The dashed lines show the difference between the four different phase-space regions. Panels (b) and (c): Excess for NATs and WATs as defined in  \autoref{eq:excess}. The dashed line represents no excess. For every phase-space region, the left marker shows the excess for tailed radio galaxies in relaxed clusters and the right the excess for tailed radio galaxies in merging clusters. The error bars are the 1$\sigma$ statistical uncertainty following \citet{2021A&A...650A.111R}.}
    \label{fig:PhaseSpaceDynState}
\end{figure*}

\begin{table*}[tbh]
\caption{Galaxy distribution in phase-space regions for all merging and relaxed clusters in the HeCS-SZ sample.}
\label{table:excessDynState}
\centering       
\begin{tabular}{c c c c c c c c c}    
\hline
Region  & $N^{Q_{i}}_{NAT,relaxed}$& $N^{Q_{i}}_{NAT,merging}$   & $N^{Q_{i}}_{WAT,relaxed}$    & $N^{Q_{i}}_{WAT,merging}$
            & $\eta^{Q_{i}}_{NAT,relaxed}$ & $\eta^{Q_{i}}_{NAT,merging}$ & $\eta^{Q_{i}}_{WAT,relaxed}$ & $\eta^{Q_{i}}_{WAT,merging}$ \\ \hline
1       & 0                        & 0                           & 0                           & 0
            & 0.00$^{+1.41}_{-0.00}$       & 0.00$^{+1.51}_{-0.00}$       & 0.00$^{+6.20}_{-0.00}$       & 0.00$^{+4.75}_{-0.00}$       \\
2       & 0                       & 4                       & 0                          & 1      
            & 0.00$^{+0.87}_{-0.00}$       & 2.19$^{+1.17}_{-0.69}$       & 0.00$^{+3.85}_{-0.00}$       & 2.19$^{+2.41}_{-0.92}$       \\
3       & 9                        & 9                           & 1                            & 1        
            & 2.22$^{+0.08}_{-0.35}$       & 1.18$^{+0.23}_{-0.26}$      & 2.22$^{+0.20}_{-1.34}$       & 0.52$^{+0.58}_{-0.22}$      \\
4       & 0                        & 3                           & 0                            & 2         
            & 0.00$^{+0.27}_{-0.00}$       & 0.51$^{+0.36}_{-0.17}$       & 0.00$^{+1.20}_{-0.00}$       & 1.37$^{+0.56}_{-0.56}$      \\ \hline
Total   & 9                        & 16                          & 0                            & 4                                     \\ \hline
\end{tabular}
\end{table*}

\begin{figure}[tbh]
\centering
\includegraphics[width=0.7\hsize]{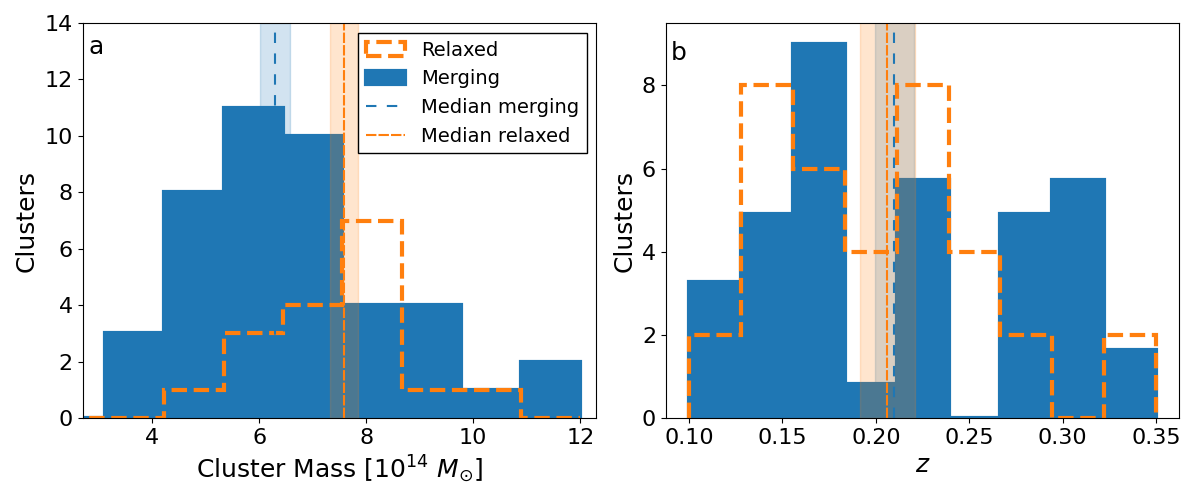}
  \caption{Distributions of the (a) cluster mass and (b) redshift for merging and relaxed galaxy clusters.}
     \label{fig:distrb_clusters}
\end{figure}

\end{appendix}

\end{document}